\documentstyle[epsf,aaspp4,flushrt,pstricks]{article}

\def\einstein	{{\em Einstein}\/}
\def\asca	{{\em ASCA}\/}
\def\bbxrt	{{\em BBXRT}\/}
\def\ginga	{{\em Ginga}\/}
\def\rosat	{{\em ROSAT}\/}
\def\bohringer	{B\"{o}hringer}
\def\muller	{M\"{u}ller}
\def\am		{$^\prime$}
\def\deg	{$^{\circ}$}
\def\kmsmpc	{~km$\;$s$^{-1}\,$Mpc$^{-1}$}

\begin{document}

\lefthead{MARKEVITCH}
\righthead{TEMPERATURE STRUCTURE OF CLUSTERS}

\title{TEMPERATURE STRUCTURE OF FOUR HOT CLUSTERS WITH \asca}

\author{Maxim~Markevitch\altaffilmark{1}}
\affil{University of Virginia, Astronomy Department, Charlottesville, VA
22903. E-mail: mlm5y@virginia.edu} 

\altaffiltext{1}{Also IKI, Moscow, Russia}

\centerline{Accepted for {\em The ApJ Letters}}

\begin{abstract}

\asca\ data are used to obtain two-dimensional gas temperature maps of the hot
non-cooling flow clusters A2256, A2319, A2163 and A665. In all four
clusters, the temperature decreases significantly at off-center distances of
$\sim 1\,h^{-1}$~Mpc ($H_0\equiv 100\,h$\kmsmpc). Central regions of the two
nearer clusters A2256 and A2319 are resolved by \asca\ and appear largely
isothermal except for the cooler spots coincident with the subunits in their
X-ray surface brightness. Although the existence of this substructure may
suggest ongoing merger activity, no asymmetric features in the temperature
distribution resembling those in the hydrodynamic merger simulations (e.g.,
Schindler \& \muller\ 1993) are apparent. In the outer parts of the
clusters, the temperature declines symmetrically with radius. In A2256 and
A2319, it follows a polytropic slope with $\gamma\simeq 1.3-1.5$. This is
somewhat steeper than the simulations predict for a flat CDM universe and is
closer to the open universe predictions (Evrard et al.\ 1996). The
temperature drop is more prominent in the outer regions of A2163 and A665
and appears even steeper than adiabatic (although not inconsistent with it).
If the gas in the outskirts of these two clusters is indeed as cool as we
measure, the cluster atmospheres should be convectively unstable and
transient. Also, such a steep temperature profile could not possibly emerge
if the gas was heated only via the release of its own gravitational energy
during infall. This may indicate the presence of an additional heat source
in the inner cluster, such as merger shocks transferring energy from the
dark matter to the gas. The results suggest that A2256 and A2319 are
pre-merger systems and A2163 and A665 are ongoing or post-mergers.

\end{abstract}


\section{INTRODUCTION}

Spatially-resolved measurements of the cluster gas temperature are necessary
for such an extensive and important problem as determining cluster masses
(e.g., Fabricant et al.\ 1984; White et al.\ 1993). Beyond that, the cluster
temperature structure can provide information on the dynamical history of
these systems. Rich and massive clusters should be just forming now in the
hierarchical clustering scenarios with a high matter density parameter
$\Omega$ (e.g., Blumenthal et al.\ 1984). On the other hand, in an open
universe most present-day clusters should be old, because their formation is
inhibited after $t\sim \Omega t_0$, $t_0$ denoting the present epoch (White
\& Rees 1978). Hydrodynamic simulations of cluster growth (e.g., Navarro et
al.\ 1995; Evrard et al.\ 1996, hereafter EMN) predict a largely constant
temperature profile in the inner part and its decline in the outer regions
for the relaxed clusters, with a steeper decline in the open universe
models. Young clusters which have recently undergone a merger should retain
a complex temperature structure (e.g., Schindler \& \muller\ 1993). However,
while there is a wealth of cluster simulations in various cosmological
scenarios, until recently, direct spatially-resolved temperature
measurements have been possible with only a limited accuracy, especially for
the hotter, more massive systems (e.g., Hughes 1991; Eyles et al.\ 1991;
Miyaji et al.\ 1993; Briel \& Henry 1994, hereafter BH; Henry \& Briel
1995). \asca\ with its broad energy coverage combined with imaging
capability (Tanaka et al.\ 1994) is set to significantly improve the
situation. Some results have already appeared (e.g., Arnaud et al.\ 1994;
Markevitch et al.\ 1994, 1996, hereafter M94 and M96; Ikebe et al.\ 1996).

In this {\em Letter}, we use \asca\ data to derive temperature maps of
nearby A2256 ($z=0.058$), A2319 ($z=0.056$), and distant A2163 ($z=0.201$)
and A665 ($z=0.18$). All four are hot, lack cooling flows and are probably
not fully relaxed, which is suggested by either the substructure in their
X-ray images or by galaxy velocities (e.g., Briel et al.\ 1991; Elbaz et
al.\ 1995). For A2256, a temperature map was earlier presented by BH who
used \rosat\ PSPC, and our results are compared with theirs. \asca\ results
on the temperature structure near the center of A2163 were reported in M94.
In M96, a steep radial temperature decrease was found in this cluster.
Interestingly, a recent measurement of the Sunyaev-Zeldovich effect toward
A2163 by Holzapfel et al.\ (1996) independently suggests a similar decrease,
although with marginal significance. Below, a less model-dependent,
two-dimensional approach to the \asca\ data is employed to confirm the
result of M96 and find similar phenomena in other three clusters.

\section{DATA AND METHOD}

For A2256, data from the two pointings were used which total 63~ks of useful
time. The pointings are offset by 6\am\ from one another, have the ``roll
angles'' between the cluster and mirror axes which differ by about 100\deg,
and are carried out with different GIS onboard background rejection modes
(high background during the early 28 ks and normal background during the
subsequent observation.) All of this facilitated useful internal consistency
checks of the results. A2319 has two pointings of 28~ks in total with a
relative offset of 12\am. For these two bigger clusters not entirely covered
by the SIS, only the GIS data were used for computational simplicity. Both
SIS and GIS were used for A2163, which has a single 27~ks pointing. A665 was
observed in one pointing for 33~ks. The detector plus sky background was
modeled using the blank field observations normalized according to their
exposures. A $1\sigma$ relative error of 20\% was assigned to these
normalizations (5\% for the A2163 GIS data, see M96). A problem was
encountered while modeling the SIS background for A665, for which the
normalization calculated this way was obviously too high.  We chose to use
only the GIS data for this cluster.

To reconstruct the cluster temperature maps, the scheme described in M96 was
used, which consists of simultaneous fitting of the spectra from all chosen
image regions, all detectors and all pointings, taking into account the PSF
scattering. The projected temperature was assumed constant within each
region of interest. The iron abundance and $N_H$ were fixed at the same
values for all regions. To minimize $\chi^2$ as a function of many free
parameters and avoid false minima, the annealing method from Press et al.\
(1992) was used. The image regions we use are all larger than 5--6\am\
across, which is sufficiently large compared to the $3'$ half-power diameter
of the \asca\ PSF. The PSF was modeled by interpolation between the GIS
images of Cyg X-1 (Takahashi et al.\ 1995). Only \asca\ data above 2.5~keV
were used due to the uncertainty of our PSF model below that energy.
Relative PSF uncertainty of 15\% ($1\sigma$) was included in the confidence
intervals calculation. For the purpose of this work, \rosat\ PSPC
images were used to model the cluster projected emission measure. They were
corrected for the gas non-isothermality in each iteration by dividing the
brightness by the plasma emissivity in the PSPC band for a given temperature
(which doesn't significantly change the results; M96). The images were
obtained using the Snowden et al.\ (1994) procedure. \rosat\ statistical and
background errors were included into the confidence intervals of our
temperature values. To correct for inaccuracy of the attitude solution,
\rosat\ and \asca\ images were aligned by eye after convolving the former
with the \asca\ response. Uncertainty of this operation (about $\pm 0.3'$ at
90\%) was included in the confidence intervals for the two big clusters.

Given the relative complexity of the method, it is useful to present some
consistency checks which were performed. Our results may be affected by the
following:

{\em Errors in the code.} Currently there are at least three other
independent techniques for the \asca\ cluster analysis, which give results
consistent with this method. An isothermal cluster was simulated by Ikebe
(1996) and it was possible to extract its input temperatures with the code
used here. R.~Mushotzky communicated that the temperature profile of A2163
derived using the code of K.~Arnaud is similar to that in M96. Using another
independent technique, Churazov et al.\ (1996) derived temperatures for
A2256 and A2319 similar to those presented here.

{\em Use of ROSAT brightness map.} An incorrect emission measure model may
result in distorted temperatures. For A2163, the temperatures and a
$\beta$-model density profile were fitted together using \asca\ only, and
the obtained density profile was in agreement with a better-constrained
\rosat\ profile (M96). For the remaining three clusters, an analogous test
was performed, in which relative normalizations between the model annuli,
set by the \rosat\ image, were freed and fitted together with the
temperatures. Their resulting values were consistent with 1, as is expected
if the \rosat\ emission measure model is adequate. Cooling flows would
require a different approach, which is the reason of our choice of
non-cooling flow clusters for the present work.

{\em PSF and effective area miscalibration.} All analysis methods are
currently using essentially the same PSF model. This model has been tested
by comparison with the point sources at different focal plane positions
(Takahashi et al.\ 1995; Ueda 1996) and found adequate to the accuracy level
which is used here. Ikebe (1996) checked the effective area calibration and
found that it is adequate after certain correction. We also note that there
are no bright sources in the vicinity of these clusters that may produce any
significant stray light contamination. For the bigger clusters A2256 and
A2319, the PSF-scattered contribution to the flux of a particular image
region, was less than a half for most of the regions, thus a PSF error does
not propagate strongly to the measured temperatures. However, images of
these clusters span the whole GIS field of view and the effective area
miscalibrations may in principle affect the results, especially for the
hotter A2319. Different pointings to these clusters help reduce the chances
of both errors. For the two smaller clusters, the PSF-scattered
contributions were about 2/3 of the outer flux and even greater for some of
the A665 regions, making the results for these clusters dependent on the
reliability of the PSF model.

It is not inconceivable that other \asca\ instrumental effects exist that
are not understood at the moment, so the best check of the results would be
that by another telescope. \rosat\ PSPC results for A2256 and A665 are
included below for this purpose. More on this comparison is presented in
Markevitch \& Vikhlinin (1996, hereafter MV).

\section{RESULTS}

For A2256, we obtained average values over the inner $r=15'$ of
$T_e=7.5\pm0.4$~keV and an iron abundance relative to Allen (1973) of
$0.23\pm0.05$\footnote{Errors are 90\% one-parameter intervals throughout.},
in agreement with \ginga\ (Hatsukade 1989). For A2319, our average
$T_e=10.0\pm0.7$~keV is in agreement with \einstein\ MPC (David et al.\
1993), and our abundance is $0.30\pm0.08$. For A665, an average temperature
of $T_e=8.0\pm1.0$~keV was found in the inner $r=10'$, in agreement with
\ginga\ (Hughes \& Tanaka 1992).  Results for the whole of A2163 were
reported in M94. For all clusters, the average spectra can be acceptably fit
with a single-temperature model. However, spatially resolved fits reveal
significant deviations from isothermality for all four clusters. Their
two-dimensional temperature maps overlaid on the \rosat\ brightness contours
are presented in Fig.~1 (Plate 1), and Fig.~2 shows projected temperatures
averaged over the concentric annuli (the A2163 profile is presented in M96).
The best-fit $\chi^2$ values for the maps in Fig.~1 are 301/408--17 d.o.f.,
220/266--14 d.o.f., 55/100--10 d.o.f.\ and 39/60--7 d.o.f.\ for A2256,
A2319, A2163 and A665, respectively, suggesting that our conservative
compound errors are, if anything, slightly overestimated. When fit
separately, different pointings to A2256 and A2319 give consistent results,
with best-fit values within the 90\% intervals for simultaneous fit in all
but a couple of regions, which is even fewer than expected from statistical
scatter.

\subsection{A2256 and A2319}

Cooler spots are observed near the centers of these two clusters, coincident
with the substructures in their X-ray images. In A2256, the approximate
subgroup region has best-fit $T_e=6.2\pm0.8$~keV, compared to an average of
8.7~keV for the immediate surrounding. It is higher than the \rosat\ value
of $3.6^{+0.9}_{-0.5}$~keV (J. P. Henry, personal comm.) and than the
\bbxrt\ value (Miyaji et al.\ 1993). Because the subgroup is projected on
the hotter main cluster, one expects that the \asca\ value would be
higher.
\footnote{For example, for a mixture of two components, an 8~keV
underlying cluster contributing 2/3 of the emission measure and a 1.5~keV
projected group contributing the rest, PSPC gives a 3.7~keV
single-temperature fit and GIS gives 6.2~keV in the 2.5--11 keV band we use.
This indicates that our results are insensitive to the cool substructure
unless it is very prominent.}
In A2319, the subgroup region has best-fit $T_e=8.4\pm1.2$~keV,
compared to an average of 10~keV at that radius.  Otherwise, the central
$r=0.5\,h^{-1}$~Mpc regions of these two clusters are roughly isothermal at
the present accuracy (the center of A2256 appears hotter than the second
ring, but only with a marginal significance). We could not confirm existence
of the two hot spots in A2256 reported in BH and ascribed to the effect of a
merger. The temperatures in our approximately correspondent regions 6 and 9
are consistent with the average temperature at this radius within the much
smaller errors (Fig.~1{\em a}). A reanalysis by MV of the \rosat\ pointings
using a scheme less sensitive to calibration uncertainties, has shown that
those hot spots are probably artifacts.

Beyond $r=0.5\,h^{-1}$~Mpc, the temperature in both clusters is found to
decline with radius, with no significant deviations from symmetry. The
\asca\ A2256 temperature profile is in good agreement with that from \rosat\
obtained by MV (Fig.~2{\em a}), while having a much better accuracy. The
\bbxrt\ measurement of Miyaji et al.\ for $r<10'$ is in agreement as
well. BH derived a different \rosat\ temperature profile (although
insignificantly so), which is addressed by MV. Using BH and Jones \& Forman
(1984) fits of the density profiles, the temperatures correspond to a
polytropic index of $\gamma\simeq 1.55$ (1.4--1.7 90\% interval) and
$\gamma\simeq 1.25$ (1.08--1.40) for outer A2256 and A2319, respectively.
For A2319, a tentative temperature estimate is also obtained for
$r\sim1\,h^{-1}$~Mpc, suggesting that the profile steepens with radius.  The
temperature profiles in these clusters are to some extent similar to that of
Coma (Hughes 1991).

\subsection{A2163 and A665}

For A2163 and A665, greater off-center distances are within the \asca\ field
of view. A2163 was reported earlier (M96) to have a largely isothermal inner
$r=0.7\,h^{-1}$~Mpc profile at around 11~keV, and beyond this, a sharp drop
of the temperature down to 4~keV at $1.5\,h^{-1}$~Mpc. An
azimuthally-resolved measurement (Fig.~1{\em c}) shows the temperature
dropping in all four directions off the cluster center, although the
individual constraints are poor.

Averaged over the central $r=0.7\,h^{-1}$~Mpc ($=6'$) region, A665 has a
temperature of $8.3\pm1.5$~keV. The temperature drops to $2.2^{+2.2}$~keV on
average in the 6--12\am\ annulus (because of our restricted energy band,
only the upper bound and not the best-fits value is meaningful.) It
decreases in all four off-center directions as well (Fig.~1{\em d}). Region
6 has a point source with an apparently non-thermal spectrum. It was fitted
simultaneously with other spectra although its influence on the other
regions was small. Since the outer cluster temperature appears to be within
the \rosat\ energy band, we have undertaken to analyze the archival 40~ks
PSPC observation of A665 to confirm the \asca\ result. The PSPC and GIS
temperatures obtained in two concentric annuli are shown in Fig.~2{\em c}.
The two instruments are in good agreement, and a temperature drop is
suggested by the PSPC data as well, although marginally significantly. The
absorption column was kept fixed at its Galactic value in the \rosat\ fit
since varying it was not required by the $F$-test, while freeing it makes
the best-fit outer temperature still lower.

The observed temperature decline in the outer part of A2163 corresponds to a
polytropic index of $\gamma\simeq 1.9$ ($\gamma>1.7$ at 90\% confidence),
while for A665, $\gamma\simeq 1.7$ ($\gamma>1.3$), adopting the density
profiles from Elbaz et al.\ (1995) and the \rosat\ image, respectively.

\section{DISCUSSION}

Although A2256 and A2319 clearly exhibit substructure in their \rosat\ X-ray
images, no large-scale merger signatures, such as those predicted by
hydrodynamic simulations (e.g., Schindler \& \muller 1993; EMN), are seen in
the temperature maps of their central parts. This may indicate that the
current mergers have not proceeded far enough to disturb the bulk of gas.
For example, Roettiger et al.\ (1995) specifically simulated A2256 and found
that the cluster image, galactic velocities and absence of the cluster-scale
temperature variations are consistent with an epoch of about 0.2~Gyr before
core passage of an infalling subunit. The observed relative symmetry of the
temperatures in A2256 and probably A2319 suggests that their outer parts
have been undisturbed by major mergers for the past few Gyr, making these
clusters good candidates for an accurate mass measurement, which will be
made in a future paper.

Apart from the subgroups, the temperature profiles of these two clusters are
qualitatively similar to those predicted by the simulations of Navarro et
al.\ and EMN for clusters in equilibrium. Interestingly, the observed
temperature decline starts at smaller radii than EMN predict for the flat
universe models without galactic winds, and is closer to their open universe
model, in which clusters are expected to have steeper density and
temperature profiles (Hoffman \& Shaham 1985; Crone et al.\ 1994; Jing et
al.\ 1995).  However, the published simulations including gas are limited to
the CDM initial perturbations spectrum, and our sample is limited to just a
couple of rather specific clusters (e.g., lacking cooling flows unlike most
of the clusters). A study of several other clusters is underway with \asca,
which will show how common this phenomenon is.

The temperature falls even steeper in A2163 and perhaps in A665. However, as
was noted in M96, the low outer values may in fact not be representative of
the mean gas temperature at those radii. For example, the measured electron
temperature may be lower than that of ions heated by shock waves, because
the timescale of electron-ion equipartition via collisions becomes
non-negligible at such low plasma densities. Other possibilities involve
cold gas clumps or point sources which cannot be localized by either \asca\
or \rosat\ but significantly contribute to the flux. On the other hand, if
the outer gas temperatures are indeed as low as measured, the observed steep
profiles would have interesting implications for the physical conditions of
the gas (although it may be premature to speculate using such poor data
constraints.) Firstly, the outer cluster parts with a steeper than adiabatic
temperature decline should be convectively unstable, and convection should
develop on a timescale of the order of the free-fall time (a few Gyr) and
erase the gradient. Thus, existence of a steep gradient implies that the
cluster cannot have remained in its present state for a longer time than
this. Another interesting problem is how such a temperature distribution may
have emerged. Early simulations of infall of the cold gas into the cluster
and its heating via the release of its potential energy (e.g., Bertschinger
1985) predict that such a process should form shallower temperature
distributions. A steeper slope may therefore indicate that the gas in the
central part has accumulated additional energy from another source.
Hydrodynamic merger simulations predict (Pearce et al.\ 1994) that during a
merger, energy is transferred from the dark matter to the gas, increasing
its entropy in the center. Thus the observed profiles may independently
indicate that these clusters have experienced major mergers.  There is some
evidence of the asymmetric temperature variations in the central part of
A2163 (M94) and recent weak lensing analysis reveals two mass peaks near its
center (Squires et al.\ 1996). A665 has a markedly asymmetric X-ray image
which may remain from a merger. Schindler \& \muller\ (1993) predict that a
merger shock wave would manifest itself at certain stages as a sharp
projected temperature gradient in the cluster outer part, not accompanied by
a similarly noticeable feature in the wide-band surface brightness, which is
what we may be observing in the two more distant clusters.

\acknowledgments

The author thanks \asca\ team for continuous support. He is grateful to C.
L. Sarazin, A. C. Fabian, R. F. Mushotzky and the referees for useful
discussions, and to A. Vikhlinin for help with the \rosat\ data analysis.
He thanks ISAS, where most of this work was done, for its hospitality and
support. Further support was provided by the Smithsonian Institution, and by
NASA grants NAG5-2526 and NAG5-1891.

\pagestyle{empty}
\clearpage

\pspicture(0,0)(20,20)
\rput[tl]{90}(0,-1){
\begin{minipage}{22cm}
{\centering\hbox to \textwidth{
\epsfysize=9cm
\epsffile{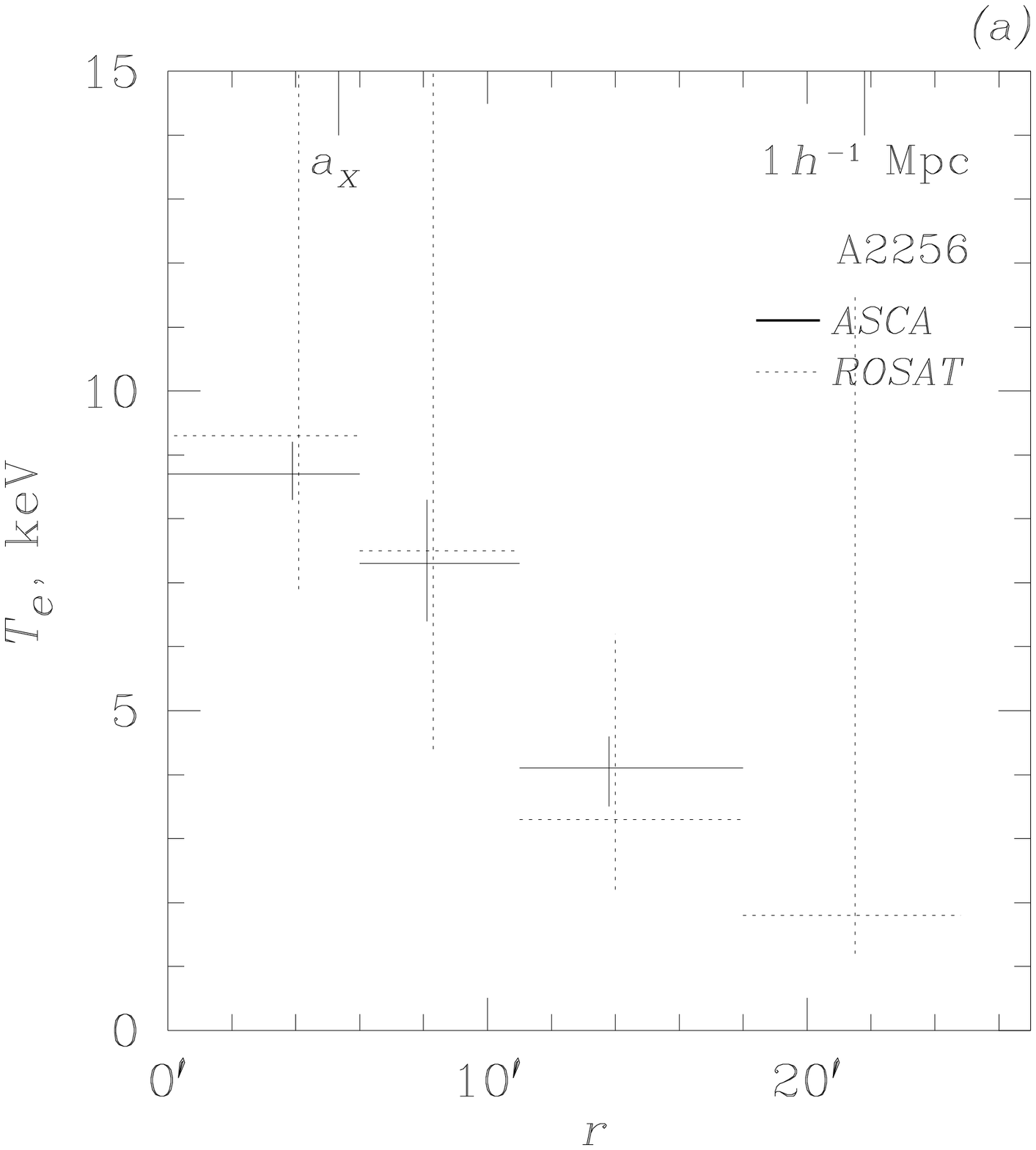} \hfill
\epsfysize=9cm
\epsffile{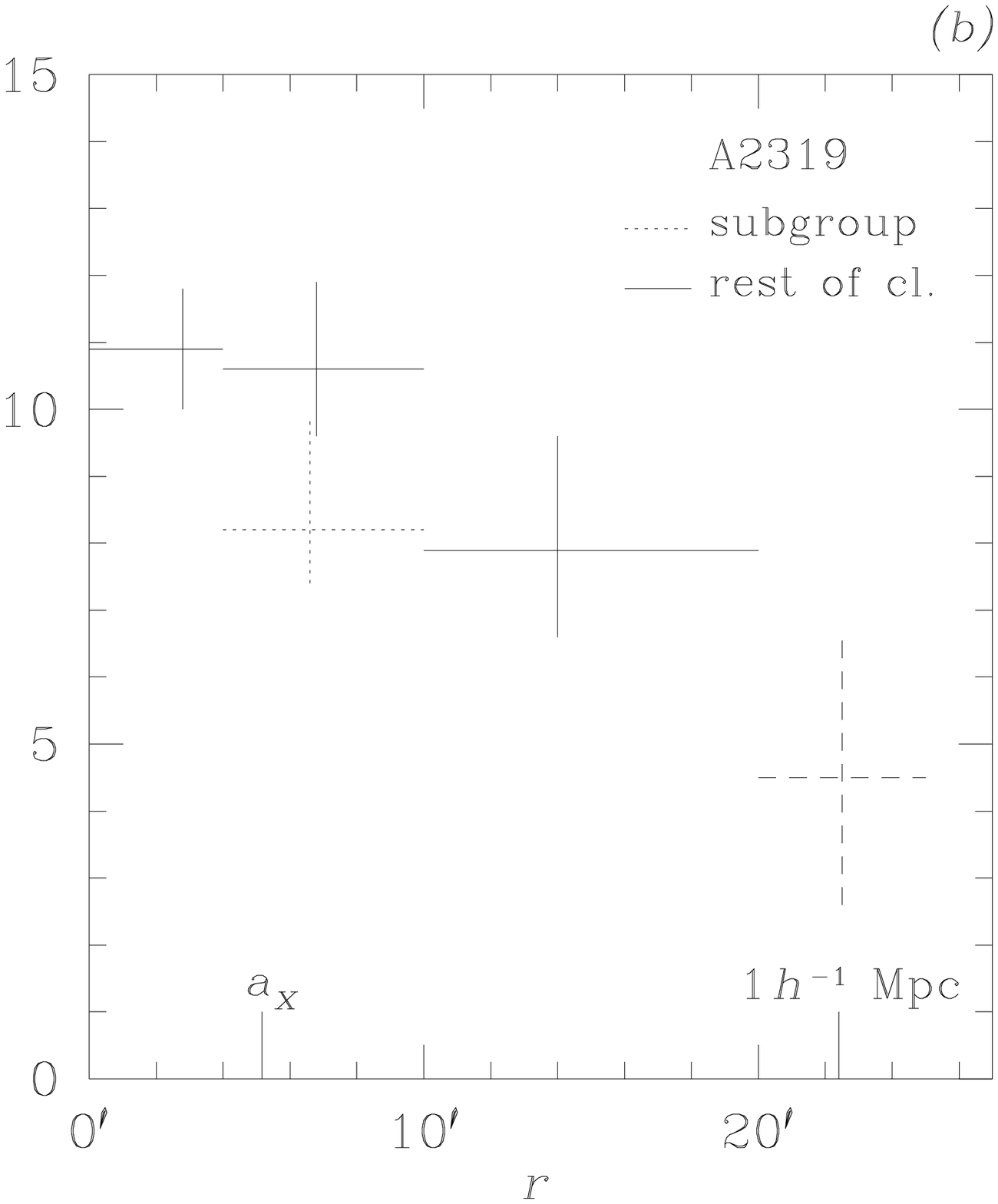} \hfill
\epsfysize=9cm
\epsffile{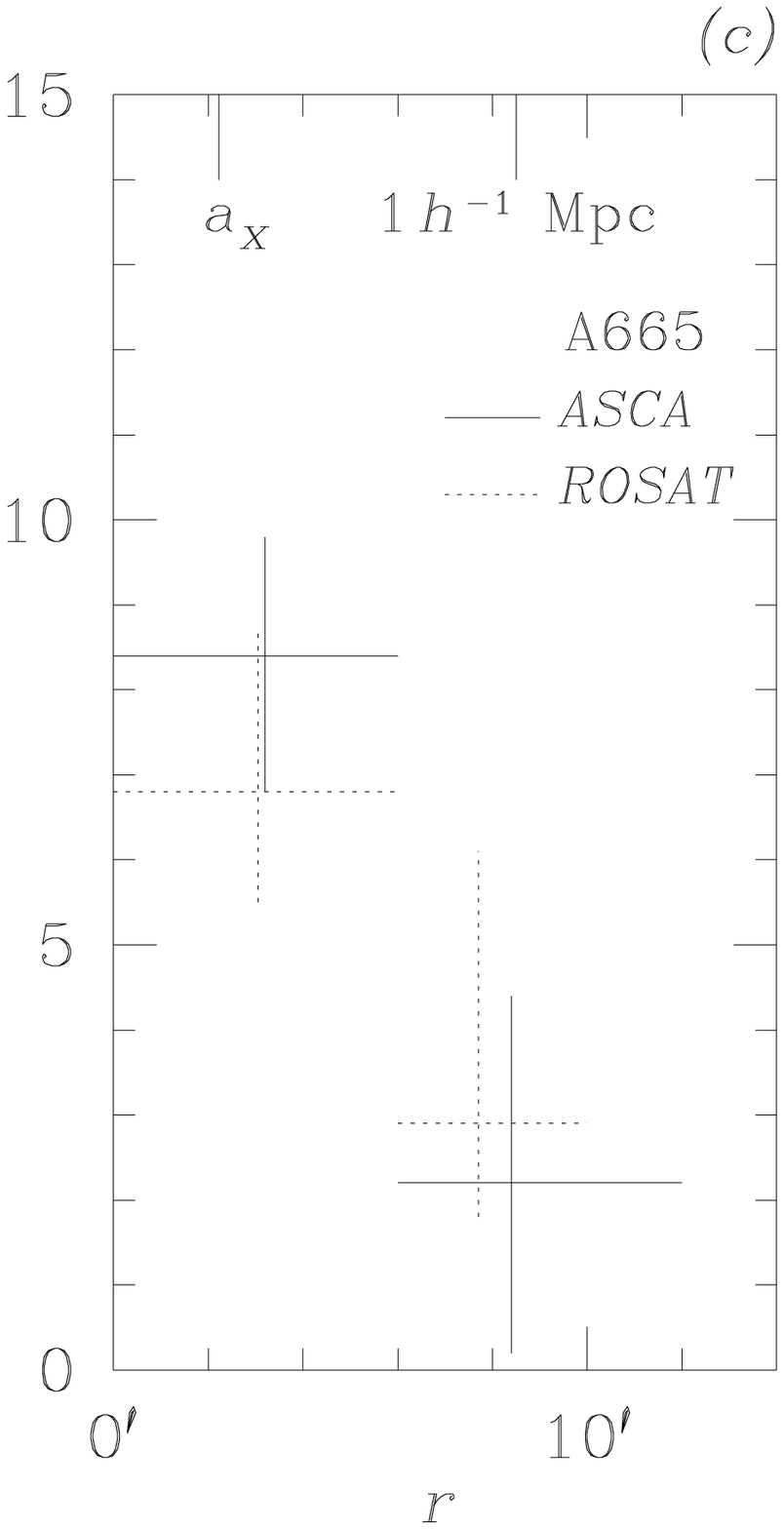}
}}
{\sc Fig.}~2.---Projected temperatures of A2256, A2319 and A665 in
concentric annuli. Errors are 90\%. Panels {\em (a)} and {\em (c)} also show
\rosat\ PSPC results from MV and this work, respectively. Parts of the
cluster affected by cooler structures in A2256 (regions 4, 9 and 10 in
Fig.~1{\em a}) and A2319 (region 2) are excluded and for A2319 shown
separately. The outer measurement in A2319, shown by dashed cross, is less
reliable because in different pointings this annulus is always close to the
edge of the FOV and covers non-overlapping parts of the cluster.
\end{minipage}
}
\endpspicture

\pspicture(0,0)(17,23)
%\psgrid(0,0)(17,23)

\rput[tl]{0}(-2,24.5){
\begin{minipage}{10cm}
\centerline{\epsfxsize=10cm
\epsffile{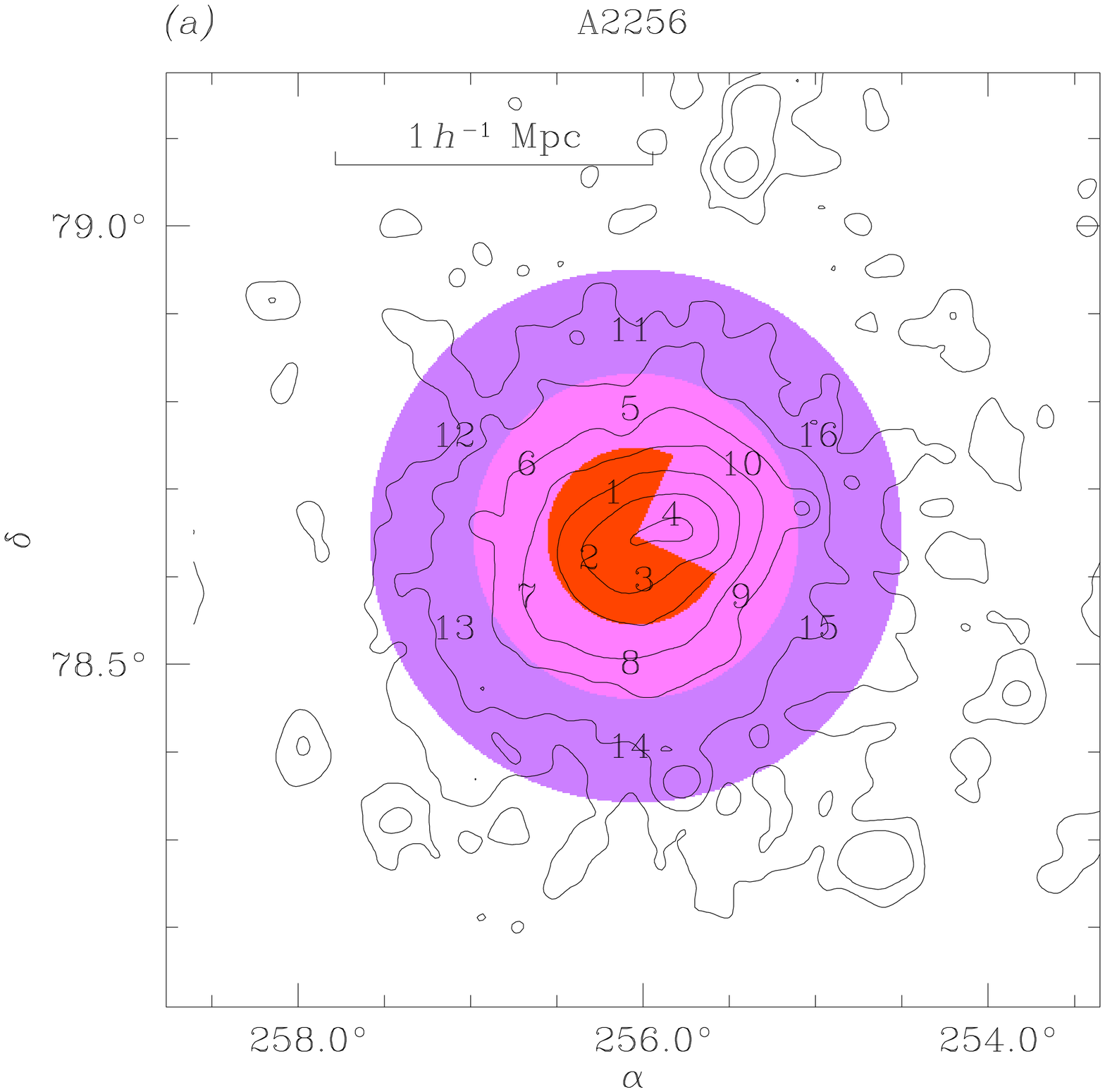}}
\vspace{-3.7mm}
\centerline{\epsfxsize=10.5cm
\hspace{4mm} \epsffile{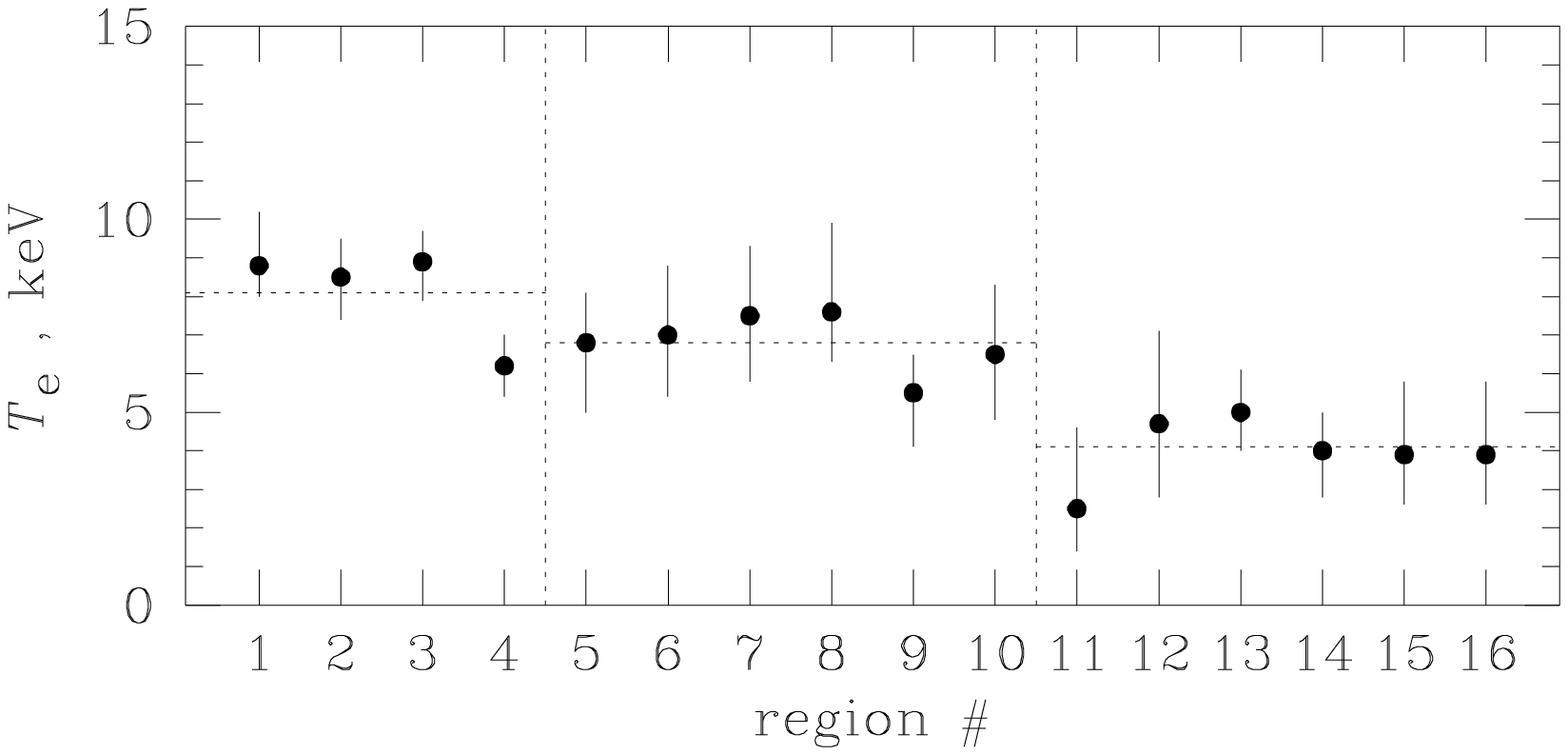}}
\end{minipage}
}

\rput[tl]{0}(7.5,24.5){
\begin{minipage}{10cm}
\centerline{\epsfxsize=10cm
\epsffile{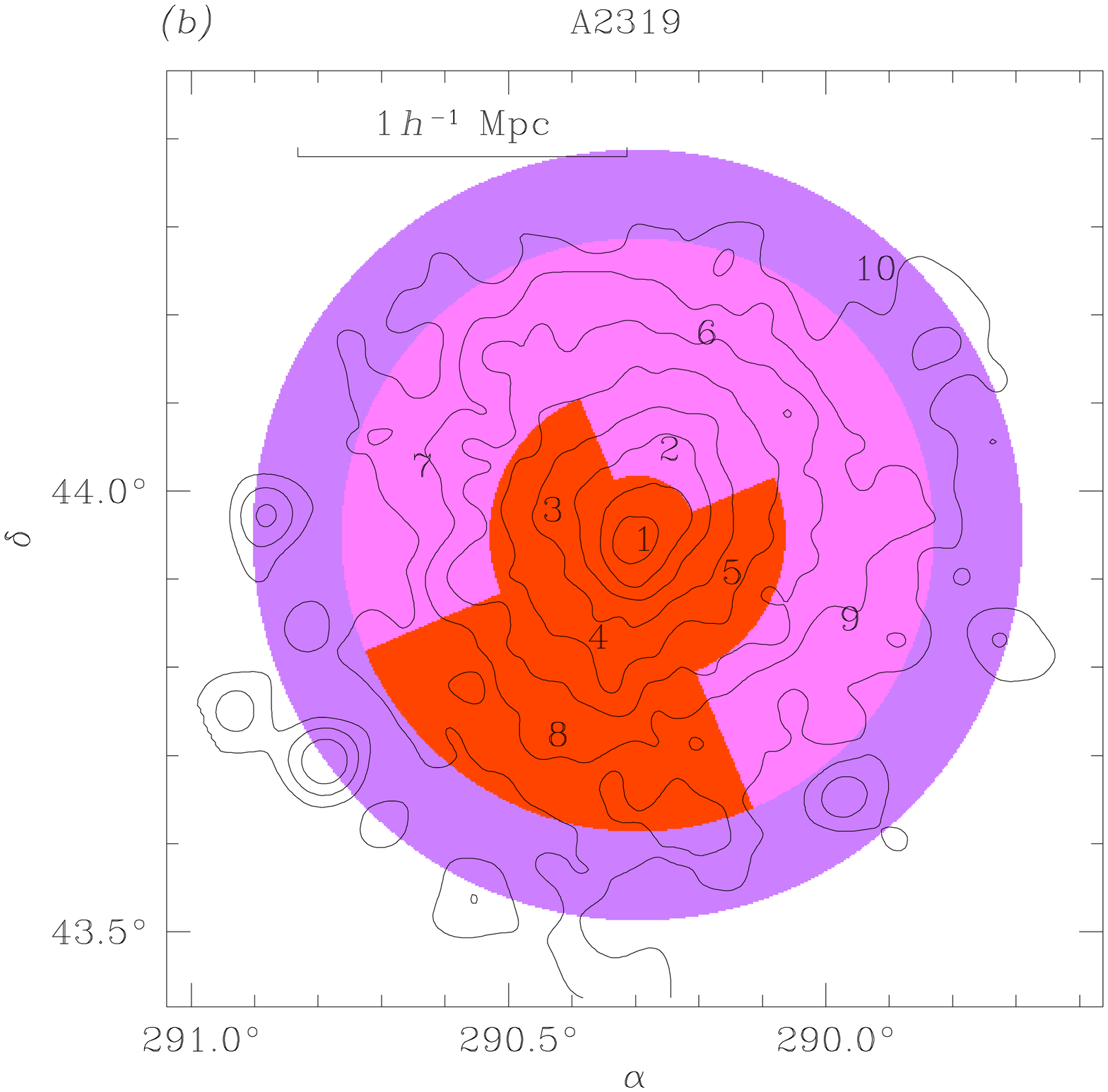}}
\vspace{-3.7mm}
\centerline{\epsfxsize=10.5cm
\hspace{4mm} \epsffile{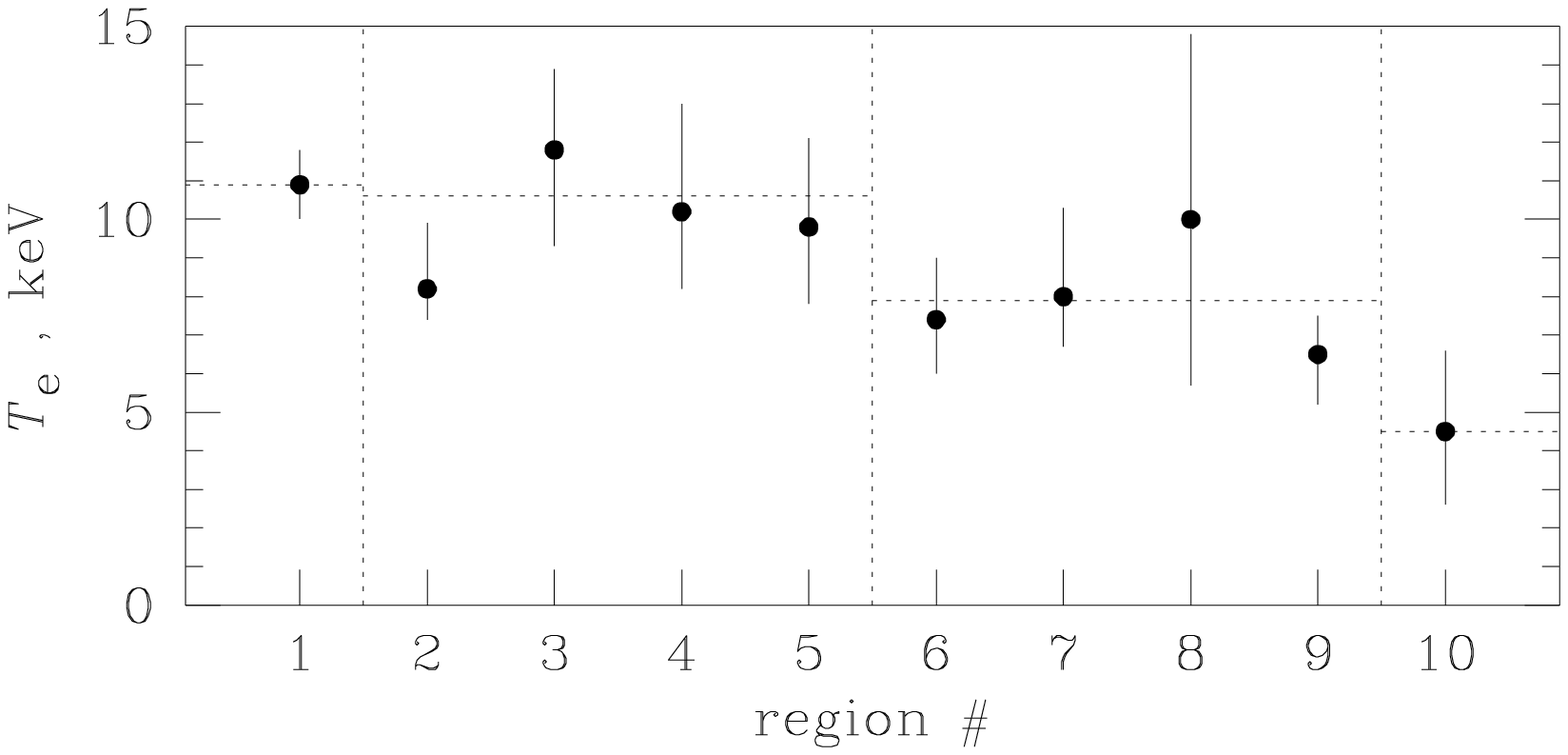}}
\end{minipage}
}

\rput[tl]{0}(-2,10){
\begin{minipage}{6.8cm}
\epsfxsize=6.8cm
\epsffile{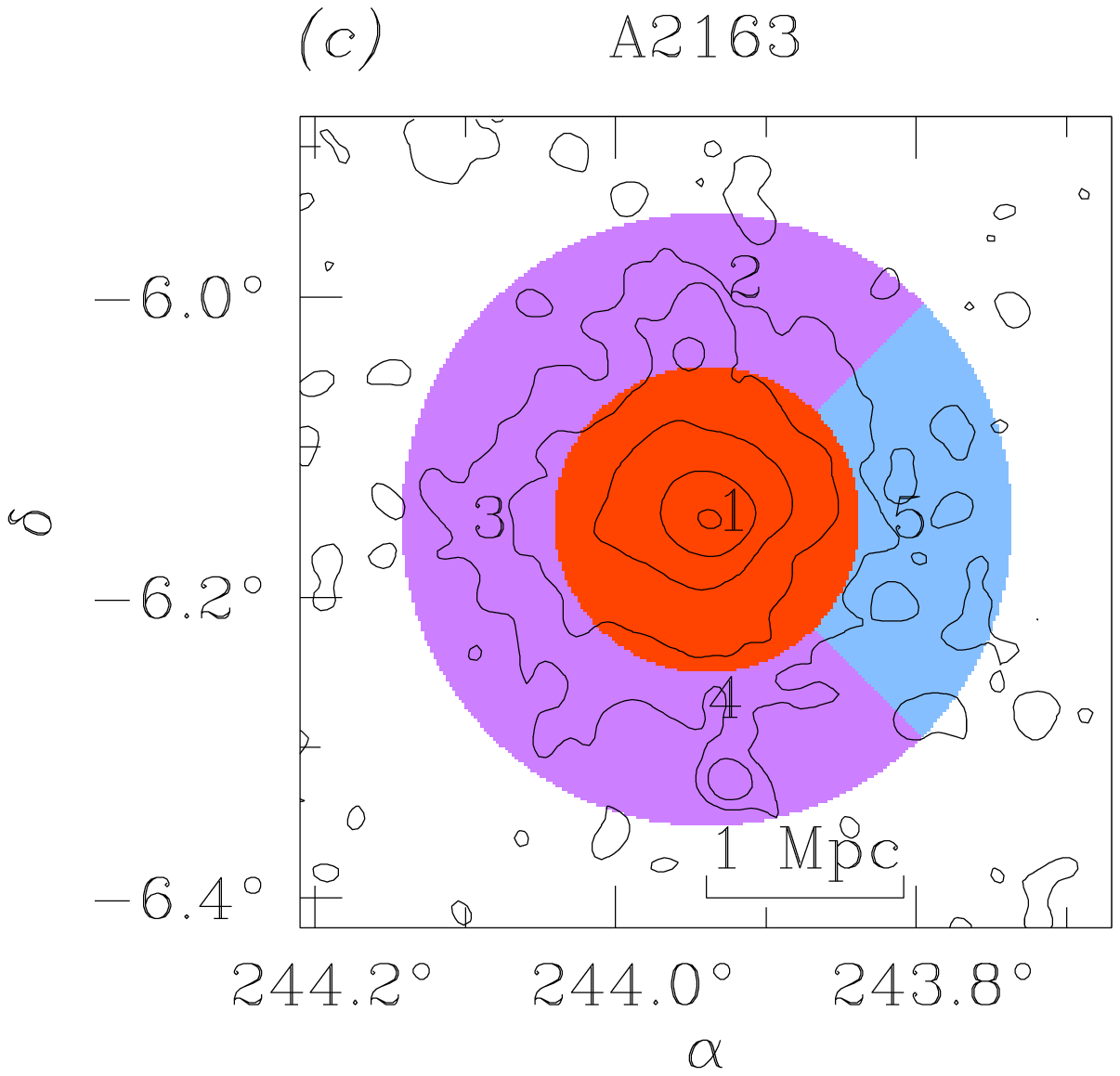}
\end{minipage}
\begin{minipage}{6.8cm}
\epsfxsize=5cm
\vspace{10.5mm}
\hspace{-17mm} \epsffile{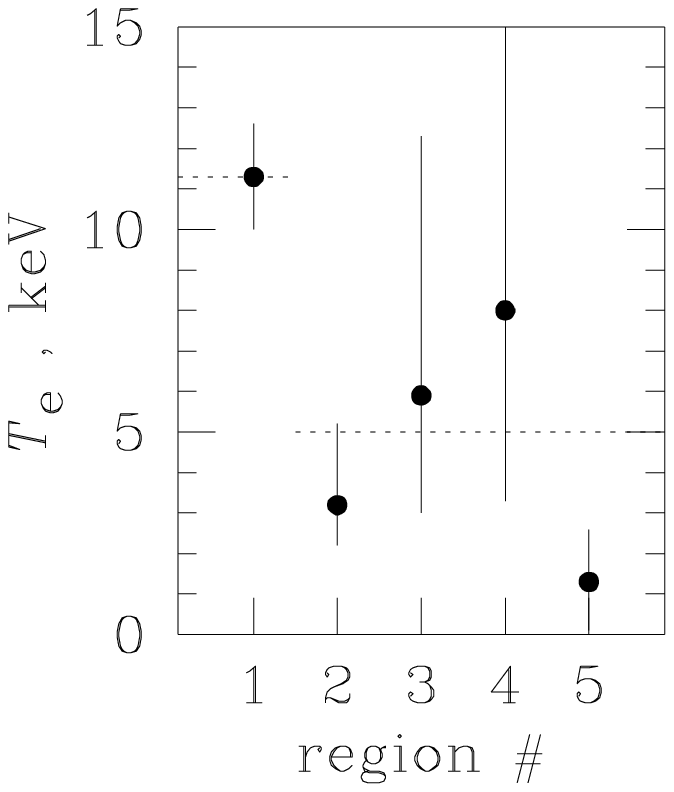}
\end{minipage}
}

\rput[tl]{0}(7.5,10){
\begin{minipage}{6.8cm}
\epsfxsize=6.8cm
\epsffile{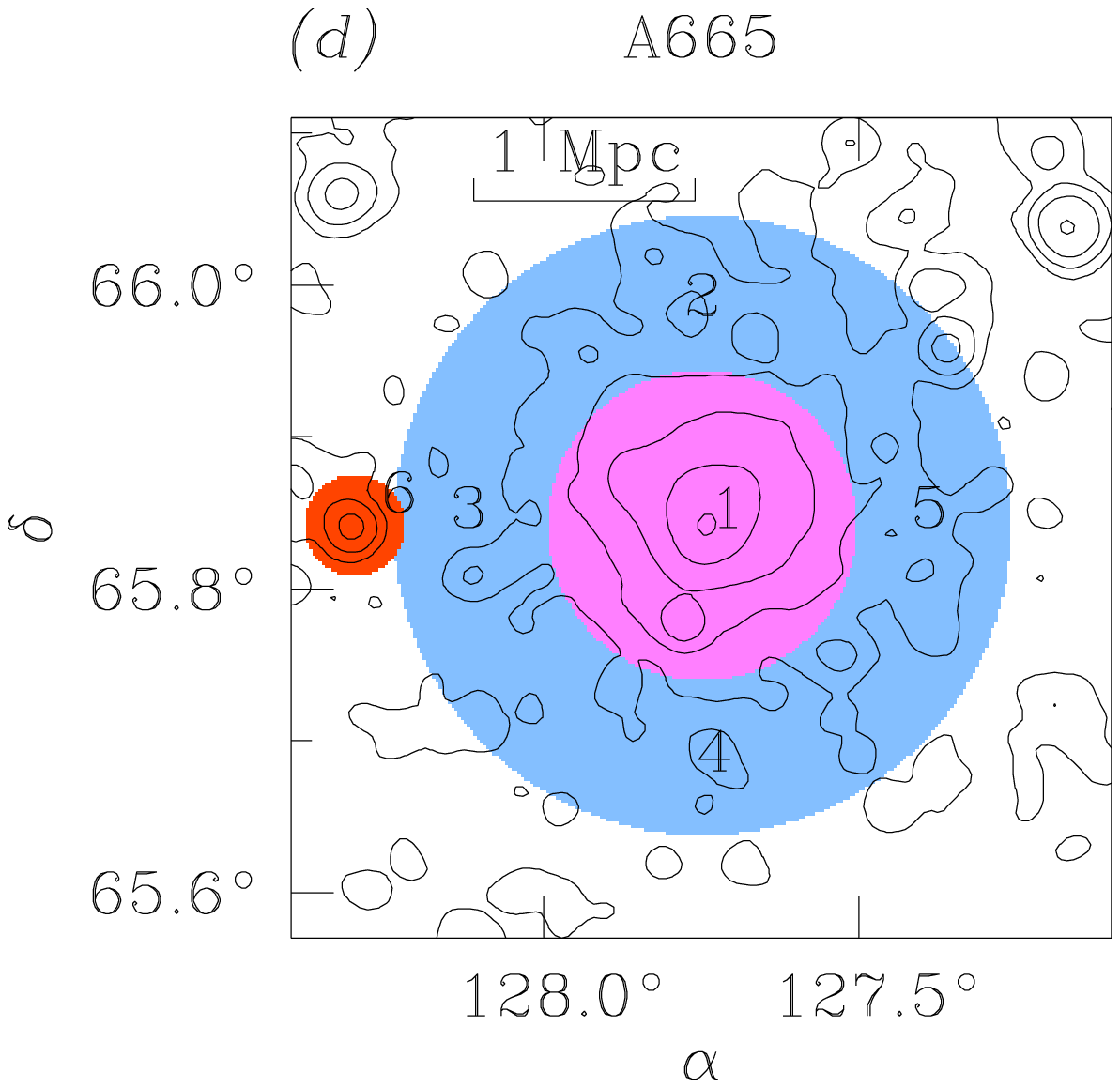}
\end{minipage}
\begin{minipage}{6.8cm}
\epsfxsize=5cm
\vspace{10.5mm}
\hspace{-17mm} \epsffile{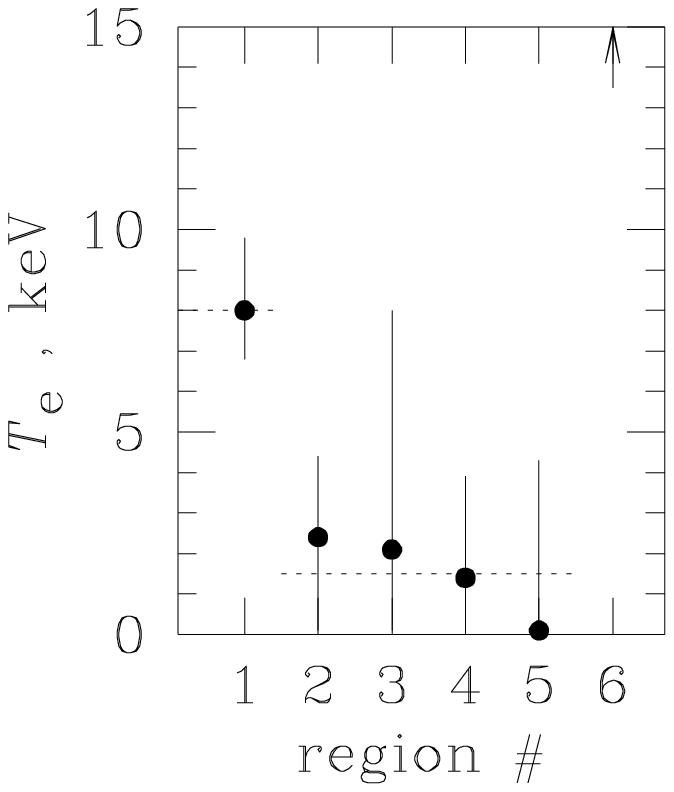}
\end{minipage}
}

\rput[tl]{0}(-1.5,4){
\begin{minipage}{18cm}
{\sc Fig.}~1.---\asca\ temperature maps (color) of A2256, A2319, A2163 and
A665, overlaid on the \rosat\ PSPC brightness contours. Regions in which the
temperature is measured are sectors of the concentric annuli. For A2256,
the regions are centered on the main subcluster peak, and for other
clusters, on the brightness peak. The regions are numbered in the maps and
their temperatures are shown in the accompanying panels with 90\% errors.
Dotted horizontal lines in those panels correspond to the average
temperature within the annulus. Regions 1 and 14 of A2319 and region 1 of
A2163 and A665 are whole rings. The outer region of A2319 is not fully
covered by the two pointings. The linear scale shown is for $h=1$.
\end{minipage}
}

\endpspicture

\end{document}